\documentclass{article}
\title{The \(\kappa\)-Minkowski Spacetime: Trace, Classical Limit and Uncertainty Relations}
\author{Ludwik D\k abrowski\thanks{SISSA, Via Beirut 2--4, 34151, Trieste, Italy}\phantom{k}\;and\ Gherardo Piacitelli\({}^*\)}
\usepackage{amsmath,amssymb}
\usepackage{ae,aecompl}
\newcommand{\Exp}{\text{Exp}}

\begin{document}
\maketitle
\begin{abstract}
Starting from a discussion of the concrete representations of the coordinates 
of the \(\kappa\)-Minkowski spacetime (in 1+1 dimensions, for simplicity), we 
explicitly compute the associated Weyl operators as functions of a pair 
of Schr\"odinger operators. This allows for 
explicitly computing the trace of a quantised function of spacetime. Moreover,
we show  that in the classical (i.e.\ large scale) 
limit the origin of space is a topologically 
isolated point, so that the resulting classical spacetime is disconnected. 
Finally we show that there exist states with arbitrarily sharp
simultaneous localisation in all the coordinates; in other words, an arbitrarily 
high energy density can be transferred to spacetime by means of localisation alone, 
which amounts to say that the model is not stable under localisation.
\end{abstract}

\section{Introduction}
The \(\kappa\)-Minkowski commutation relations are 
\cite{lukierski,lukierski2,majid}
\[
[q^0,q^j]=\frac i\kappa q^j,\quad [q^j,q^k]=0,\quad {q^0}^*=q^0,
\quad {q^j}^*=q^j,
\]
where \(k,j=1,\dotsc,d\).
Usually the real parameter \(\kappa\)
is taken of order of a Planck mass; here we will set \(\kappa=1\) 
(natural units). For simplicity we specialise to the case \(d=1\); defining
\(T=q^0\), \(X=q^1\), these commutation relations become
\begin{equation}
\label{eq:formal_basic_rels}
[T,X]=iX.
\end{equation}
However, our remarks take over to the general case, including the physically 
relevant \(d=3\); in this short note we outline some of 
the results of a forthcoming paper \cite{dp}.

We will begin by fixing an appropriate definition of regular 
representations, which amounts to formulate them in the stronger form of commutation
relations between the unitary one parameter groups generated by \(X\) and \(T\).
Accordingly, we will describe all the irreducible regular representations 
(see also \cite{agostini}).
We will observe that, since the spectrum \(\mathbb R\) of the 
most general regular position operator
\(X\) is singular in the origin for any value of \(\kappa\), the origin
will remain an isolated point also in the classical limit. In other words,
in the classical limit of the \(\kappa\)-Minkowski spacetime it will 
not be allowed to continuously travel from one side to the other of the origin:
an impenetrable barrier will cut the limiting classical spacetime 
in two decoupled halves.

We then will compute 
explicitly the Weyl operator \(W(\alpha,\beta)=e^{i(\alpha T+\beta X)}\). 
The composition 
rule of Weyl operators which results from the regular commutation relations matches
those obtained by integrating the Baker--Campbell--Hausdorff (BCH)
formula \cite{lukierski_BCH},
thus justifying a posteriori the application of abstract Lie algebraic methods
to the calculus with the Weyl operators.
Correspondingly, we will obtain a well defined 
star product associated with the Weyl calculus 
(see also \cite{lizzi,lizzi_f}). 
In addition, the explicit knowledge of the Weyl operators
will enable us 
to explicitly compute the trace of the associated Weyl 
quantisation of a classical function. In the appendix, we briefly recall 
the relationship between commutation relations of 
operators and Lie type relations, in order to ease the comparison with 
the existing literature.

We will complement our discussion by describing how to provide states saturating
the Heisenberg uncertainty relations implied by the commutation relations.
Indeed, Heisenberg theorem only provides us with lower bounds.

At the end we draw some conclusions. 

\section{Representations and Classical Limit}
\label{sec:repr}

The relations \eqref{eq:formal_basic_rels} are not sufficient to fix a unique 
model, and we need a regular form (see the appendix for a reminder 
of motivations). In order to guess it, 
we first seek for a nontrivial representation (\(X\neq 0\)). Using the well 
known relation \([P,f(Q)]=-if'(Q)\), where \(P=-id/ds,Q=s\cdot\) 
are the usual Schr\"odinger operators on \(L^2(\mathbb R,ds)\), we easily find
a representation by setting
\[
T=P,\quad X=e^{-Q}.
\]

By computing the explicit action of the unitary groups 
\(e^{i\lambda T},e^{i\lambda X}\), we find
\begin{equation}\label{eq:def_reg_rep}
e^{i\alpha T}e^{i\beta X}=e^{i\beta e^{-\alpha}X}e^{i\alpha T},\quad \alpha ,\beta \in\mathbb R,
\end{equation}
Operators \(T,X\) fulfilling the above relations
are said---by definition---a~regular 
representation of the relations \eqref{eq:formal_basic_rels}.

It is now immediate to check that the choice \(T=P,X=-e^{-Q}\) 
(note the sign) also fulfils
the above relations; indeed it can be shown directly \cite{dp} 
that the two pairs \((T=P,X=\pm e^{-Q})\) are the only irreducible, non trivial 
representations of the relations \eqref{eq:def_reg_rep}, up to unitary
equivalence. The uniqueness 
argument relies essentially on that of von Neumann for the Schr\"odinger 
operators. Our results are essentially equivalent to those of \cite{agostini}, 
obtained in a different setting using the theory of induced representations.

The trivial representations\footnote{by Schur's 
lemma, the irreducible trivial representations act on the one dimensional
Hilbert space \(\mathbb C\), and \(T=c\cdot\) is a multiple of the identity
with \(c\in\mathbb R\); standard direct integral techniques yield 
a highly reducible trivial representation which contains all the trivial ones 
precisely once: the latter can be equivalently obtained by setting
\(T=Q,X=0\) on \(L^2(\mathbb R)\).} are, by definition, 
those where \(T\) is any selfadjoint operator, and \(X=0\).

To sum up, for an irreducible regular representation, there are only three 
possibilities:
\begin{enumerate}
\item \(X\) is strictly positive;
\item \(X\) is strictly negative;
\item \(X=0\).
\end{enumerate} 

By definition, the most general admissible representation \(T,X\) 
of our relations will have to
contain all the above mentioned irreducibles, otherwise the position operator
\(X\) would fail to have the whole \(\mathbb R\) as its spectrum, and we would
not be entitled to call our model a quantisation of \(\mathbb R^2\). In what 
follows, we shall term ``universal'' the representation which contains 
any irreducible precisely once; it generates the C*-algebra of the commutation
relations, which turns out to be \(\mathcal K\oplus\mathcal C_0(\mathbb R)\oplus
\mathcal K\), where \(\mathcal K\) is the algebra of compact operators on the separable,
infinite dimensional Hilbert space.

More precisely, the most general admissible position operator
\(X\) has spectrum \(\sigma(X)=\mathbb R\) decomposed as
\[
\sigma(X)=\sigma_{\text{sing}}(X)\cup\sigma_{\text{cont}}(X),
\]
where the singular and continuous spectra are
\[
\sigma_{\text{sing}}(X)=\{0\},\quad \sigma_{\text{cont}}(X)=\mathbb R-\{0\}.
\] 
 
Now, it is remarkable that the above does not depend on the value
of \(\kappa\) (here set equal to one), hence it is bound to survive 
the classical limit. This means that, as \(\kappa\rightarrow\infty\),
\(X\) will go to a continuous function (the usual coordinate function \(x\))
of \(\mathbb R\), which will have \(0\) as an isolated point of its  
range\footnote{Note that, in 
an algebra of continuous functions, the range is the same as the spectrum.}. 
For this not to be in conflict with the asserted continuity, 
\(\mathbb R\) must come equipped with an unusual topology which makes \(0\) 
an isolated point. Thus, the classical limit of the two dimensional
\(\kappa\)-Minkowski spacetime is \(\mathbb R\times\mathbb R\) as a set; but 
as a topological space, it equals  
\(\mathbb R\times\tilde{\mathbb R}\)  where
\[
\tilde{\mathbb R}=(-\infty,0)\sqcup\{0\}\sqcup(0,\infty)
\]
is the topologically disjoint union of the two open half lines and the origin.

\section{Weyl Operators and Quantisation}
\label{sec:weyl_ops}

A direct, explicit computation of the Weyl operators 
\(W(\alpha,\beta)=e^{i(\alpha T+\beta X)}\) is not an easy task. It is much 
easier to guess them, and check the guess a posteriori by means of
the Stone--von Neumann theorem. To this end, 
we remark that the operators we seek for
should fulfil three evident requests:
\begin{gather*}
W(\alpha ,0)=e^{i\alpha T},\quad W(0,\beta )=e^{i\beta X},\\
W(\alpha ,\beta )^{-1}=W(\alpha ,\beta )^*,\\
W(\lambda \alpha ,\lambda \beta )W({\lambda'} \alpha ,{\lambda'} \beta )=
W((\lambda+{\lambda'})\alpha ,(\lambda+{\lambda'}) \beta )
\end{gather*}
identically for \(\alpha ,\beta ,\lambda,{\lambda'}\in\mathbb R\).
The last requirement expresses the remark that, for each \(\alpha,\beta\)
fixed, the operator \(\alpha T+\beta X\) is selfadjoint, so that 
\(\lambda\mapsto W(\lambda\alpha,\lambda\beta)=e^{i\lambda(\alpha T+\beta X)}\) 
is a unitary one parameter group.
With the ansatz \(W(\alpha,\beta)=e^{ir(\alpha,\beta)T}e^{is(\alpha,\beta)X}\), 
some little effort \cite{dp} leads to the solution
\[
W(\alpha,\beta)=
e^{i\alpha T}e^{i\frac{e^{\alpha} -1}{\alpha }\beta X}.
\]
In particular if \(T=P,X=\pm e^{-Q}\), then we have
\begin{equation}\label{eq:expl_weyl} 
(W(\alpha,\beta)\xi)(s)=(e^{i\alpha P\pm\beta e^{-Q}}\xi)(s)=
e^{\pm\frac{1-e^{-\alpha}}{\alpha}\beta e^{-s}}\xi(s+\alpha),\quad\xi\in L^2(\mathbb R).
\end{equation}

It is now a routine check to see that, due to the commutation relations, the product of two Weyl operators is again a Weyl operator:
\[
W(\alpha_1,\beta_2)W(\alpha_2,\beta_2)=W(\alpha,\beta),
\]
where  
\[
(\alpha,\beta)=
(\alpha_1+\alpha_2,
w(\alpha_1+\alpha_2,\alpha_1)e^{\alpha_2}\beta_1
+w(\alpha_1+\alpha_2,\alpha_2)\beta_2)\]
is defined in terms of the function
\[
w(\alpha,\alpha')=\frac{\alpha(e^{\alpha'}-1)}%
{\alpha'(e^{\alpha}-1)}
\]
which is understood to be extended by continuity to the whole \(\mathbb R^2\).

In other words, the set of Weyl operators is a subgroup of the group
of unitary operators. Moreover, since for the universal representation \((T,X)\)
the correspondence
\[
(\alpha,\beta)\leftrightarrow W(\alpha,\beta)
\]
between \(\mathbb R^2\) and the group of Weyl operators is one to one, we may 
use it to endow \(\mathbb R^2\) with a new group structure, with product
\[
(\alpha_1,\beta_1)(\alpha_2,\beta_2)=(\alpha_1+\alpha_2,
w(\alpha_1+\alpha_2,\alpha_1)e^{\alpha_2}\beta_1
+w(\alpha_1+\alpha_2,\alpha_2)\beta_2)
\]
and identity \((0,0)\). We will denote the resulting group by \(H\). This group
will play a role analogous to that played by the Heisenberg group in the case
of the canonical commutation relations (CCR) \cite{neumann}; we refrain from 
calling it the \(\kappa\)-Heisenberg 
group, because such terminology already arose in the framework of quantum 
groups; moreover, \(H\) is not a deformation of the Heisenberg group.

The group \(H\) is connected and simply connected; hence it is uniquely 
associated to its Lie algebra \(\text{Lie}(H)\) which is precisely the real
Lie algebra with generators \(u,v\) and relations \([u,v]=-v\) \cite{dp}.
 
The natural ansatz for the quantisation is to interpret the Weyl operators
as the quantised plane waves; correspondingly, for a generic ordinary 
function \(f\) we set
\[
f(T,X)=\int d\alpha d\beta\;\check f(\alpha,\beta)W(\alpha,\beta) 
\]
where
\[
\check f(\alpha,\beta)=\frac{1}{(2\pi)^2}\int dt dx\;f(t,x)
e^{-i(\alpha t+\beta x)}.
\]
The star product is then defined by
\[
(f\star g)(T,X)=f(T,X)g(T,X),
\]
where the operator product is taken on the right hand side. With this position,
a formal computation yields
\begin{eqnarray*}
\lefteqn{(f\star g)(t,x)=}\\
&=&\frac{1}{(2\pi)^2}\int d\alpha\; d\beta\; dy\; dz\;
%w(\alpha-\beta,\beta)e^{\beta-\alpha}\\
e^{i(\alpha t-\beta y-\alpha z+\beta z)}
f(y,w(\alpha,\beta)e^{\alpha-\beta}x)
g(z,w(\alpha,\alpha-\beta)x).
\end{eqnarray*}

Indeed, it is rather difficult to directly check that these definitions
are well posed for a sufficiently rich class of functions, 
closed under this product. The reason for this is that, 
contrary to the case of the Heisenberg group, our 
\(H\) is not unimodular. However
there is a way out, which will be described in detail in \cite{dp}.

\section{The Trace}
\label{sec:trace}
We now will be rewarded of the effort we spent in carrying on explicit
computations: we will classify the functions \(f\) whose quantisations 
\(f(T,X)\) are trace class as operators on a Hilbert space, 
and compute explicitly the trace of \(f(T,X)\)
by means of  a functional evaluated on the function \(f\), 
when \(T,X\) is the universal 
representation.

Let us first fix the representation 
\((T_+=P,X_+=e^{-Q})\) with positive \(X\);
given a function \(f\), we seek for a function
\(g\) such that 
\begin{equation}\label{eq:ftog}
f(T_+,X_+)=g(P,Q),
\end{equation}
where for the right hand side we take the canonical Weyl quantisation
for the Schr\"odinger particle on the line. Then we use the 
well known fact that
\[
\text{Tr}(g(P,Q))=\int dt dx\;g(t,x)=:\tau_+(f).
\]
Analogously, fixing the representation with negative \(X\), we define
\(\tau_-\). For admissible \(f\)'s, the functionals \(\tau_\pm(f)\) 
only will depend on the values \(f\) takes on the half lines \(\pm(0,\infty)\),
respectively. Finally, for the universal representation we find
\[
\text{Tr}f(T,X)=\tau_-(f)+\tau_+(f),
\]
where
\[
\tau_\pm(f)=\int dt\;dx f(t,\pm e^{-x}),
\]
and \(f(T,X)\) is trace class if and only if both integrals exists (so that
\(f(t,0)=0\) is a necessary, yet not sufficient condition for \(f(T,X)\) to 
be trace class).

To complete the above discussion, we have to show how to determine the 
functions 
\(g\). Here the idea is to compare the integral kernels \(K^f_\pm\) of
\(f(T=P,X=\pm e^{-Q})\) and \(H^{g}\) of \(g(P,Q)\), specified here below: 
for \(\xi\in L^2(\mathbb R)\),
\begin{gather*}
(f(P,\pm e^{-Q})\xi)(s)=\int dr\; K^f_\pm(s,r)\xi(r),\\
(g(P,Q)\xi)(s)=\int dr\; H^g(s,r)\xi(r),
\end{gather*}
where
\begin{gather*}
K_\pm^f(s,r)=
\frac{1}{2\pi}\int dt\;f\left(t,\pm\frac{e^{-s}-e^{-r}}{r-s}
\right)e^{i(r-s)t},\\
H^g(s,r)=\frac{1}{2\pi}\int dt\; g\left(t,\frac{r+s}{2}\right)e^{i(s-r)t}.
\end{gather*}
The kernel \(H^{g}\) is well known from canonical (CCR)
Weyl quantisation; \(K^f_\pm\)
can be directly computed using the explicit action \eqref{eq:expl_weyl} 
The condition \eqref{eq:ftog} then becomes
\[
H^{g}\equiv K^f_\pm,
\]
which has solution
\[
g_\pm(t,x)=
\int d\alpha\;e^{i\alpha t}f^{\check\ \otimes\text{id}}
\left(\alpha,\pm\frac{e^{\alpha/2}-e^{-\alpha/2}}{\alpha}e^{-x}\right).
\]
We refer the interested reader to \cite{dp} for a more detailed discussion.

\section{Uncertainty Relations}

Of course, in any trivial representation \(T,X\) commute, hence Heisenberg
uncertainty has empty content, and simultaneous sharp 
localisation can be obtained both in space and time with states relative to a trivial representation. 
This might seem specifically related to the special status of \(0\) 
in the spectrum of the most general
(universal) representation; that however is not the case, indeed. 
Fix for example the irreducible 
representation with positive \(X\), and observe that, due to the form 
\(X=e^{-Q}\) of the position, a state \(\xi\) is localised close to 0
if, as an \(L^2\) function of \(s\), it is essentially 
supported at large positive \(s\).
In other words, the behaviour of \(\xi\) at large (small) \(s\) is related 
with localisation of \(X\) at small (resp. large) spectral values of 
\(X=e^{-Q}\). Hence one may take a state with any desired 
uncertainty \(\varepsilon\) in \(T=P\); such a state can be chosen with finite
(though large) support as a function of \(s\). By shifting it 
(as a function of \(s\)) on the right sufficiently far from \(s=0\),
one may obtain a state sharply localised around the spectral value \(0\) of 
\(X=e^{-Q}\) with any given uncertainty \(\eta>0\), without affecting
the uncertainty \(\varepsilon\). Hence the two uncertainties 
\(\varepsilon\) and \(\eta\)  can be chosen independently, and small
at wish. 

In conclusion, it is possible to simultaneously localise in time and space,
at the only cost of confining the state sufficiently close to the origin. 
One might give an intuitive description of this state of affairs by saying that
the \(\kappa\)-Minkowski spacetime is classical (at any time) 
close to the origin of space; while, by similar arguments, one might say that
it is increasingly noncommutative far away from the origin of space (e.g. at cosmic distances from the origin). Note that,
together with the breakdown of translation covariance 
(implicit in the commutation 
relations), this gives the origin of space a very special status.

\section{Conclusions}
While a thorough 
mathematical discussion of the representation theory (and thus of the
associated C*-algebra) is available, giving a complete symbolic calculus
in terms of star products and traces associated to the quantisation
\`a la Weyl, on the other side the physical interpretation of the model 
exhibit some unpleasant features. In particular,
the classical limit, though describing the usual spacetime as a set, appears to
be endowed with a pathological topology. Moreover, contrary to any physical 
expectation, it exhibits very large noncommutative effects at large (e.g. 
cosmic) 
distances from a privileged point of the space. These features become even 
more strikingly 
unpleasant in higher dimensions  \cite{dp}.

\section*{Appendix}

We recall here some 
basic facts about representations of Lie relations by selfadjoint
operators on a Hilbert space. Firstly, we wish to 
fix the correspondence 
between the abstract real Lie relations and the commutation
relations of the associated regular representations, which involve the 
complex structure. Secondly, we wish to emphasise in general that the real Lie 
algebra underlying the definition of regular representations of  
the given commutation relations among Hilbert space operators plays
an ancillary r\^ole.

Let \(\mathcal A\) be a real Lie algebra with generators 
\(u_1,\dotsc,u_n\) and relations
\begin{equation}\label{eq:abstr_lie}
[u_j,u_k]=\sum_lc_{jkl}u_l,
\end{equation}
and consider a representation \(U\) 
(by unitary operators on some Hilbert
space \(\mathfrak H\)) of the unique connected, simply
connected group \(G\) with \(\text{Lie}(G)=\mathcal A\). 
If \(\Exp:\mathcal A\rightarrow G\) is the usual Lie exponential map, 
there are uniquely defined selfadjoint operators \(A_1,\dotsc,A_n\) on 
\(\mathfrak H\), such that 
\(U(\Exp[\lambda u_j])=e^{i\lambda A_j}\) as unitary one--parameter 
groups of operators.
For every choice of the generators \(u_{j_1},\dotsc,u_{j_k}\)
there are \(\mathcal A\)-valued functions
\(\alpha^{(k)}_{j_1,\dotsc,j_k}\) defined on some 
open neighbourhood of the origin
of \(\mathbb R^k\) such that
\[
\Exp[\lambda_1 u_{j_1}]\Exp[\lambda_2 u_{j_2}]\dotsm
\Exp[\lambda_k u_{j_k}]=\Exp\left[
\sum_l\alpha^{(k)}_{j_1,\dotsc,j_k}(\lambda_1,\dotsc,\lambda_k)\right].
\]
Correspondingly, there are selfadjoint operator valued functions 
\(R^{(k)}_{j_1,\dotsc,j_k}\) such that
\begin{equation}
\label{eq:gen_weyl_rel}
e^{i\lambda_1 A_{j_1}}e^{i\lambda_2 A_{j_2}}\dotsm e^{i\lambda_k A_{j_k}}
=e^{iR^{(k)}_{j_1,\dotsc,j_k}(\lambda_1,\dotsc,\lambda_k)}.
\end{equation}
Formal computations yield
\[
\left.
\frac{d^2}{d\lambda\;d\lambda'}e^{i\lambda A_j}e^{i\lambda' A_k}e^{-i\lambda A_j}
\right|_{\lambda=\lambda'=0}=i[A_j,A_k].
\]
Hence, using \eqref{eq:gen_weyl_rel}, we get
\begin{equation}
\label{eq:gen_comm_rel}
[A_j,A_k]=i C_{jk},
\end{equation}
where 
\[
C_{jk}=
-\left.
\frac{d^2}{d\lambda\;d\lambda'}R^{(3)}_{j,k,j}(\lambda,\lambda',-\lambda)
\right|_{\lambda=\lambda'=0}.
\]

The commutation relations \eqref{eq:gen_weyl_rel} are usually called a 
regular (or Weyl) form of the commutation relations \eqref{eq:gen_comm_rel}, relative to the given representation \(U\) of \(G\). In order to give 
them an intrinsic meaning, one has to give a criterion to select \(G\) and
\(U\).
Typically, the fundamental physical 
relations are those in the ordinary form \eqref{eq:gen_comm_rel}
(to be complemented with the implicit requirement that 
\(A_j=A_j^*,C_{jk}=C_{jk}^*\)), which
are directly related to physical interpretation through the Heisenberg theorem.
The {\itshape choice} 
of the corresponding regular relations (i.e.\ of \(G\) and \(U\))
is then a subsequent step which is 
necessary in order to fix the admissible realisations of the model. 

The many technical problems afflicting the above formal derivation
of \eqref{eq:gen_comm_rel} (as well as its interpretation) should not be 
considered just as ``technicalities'' of no physical interest: 
indeed, even when
the basic  
relations \eqref{eq:gen_comm_rel} are nicely fulfilled on some dense domain,
they may belong to the representation of a totally different Lie 
algebra, namely to totally different Weyl relations.
There is a striking example, 
due to Nelson (unpublished; see \cite[VIII.5]{rs}), 
of two operators which are essentially selfadjoint 
on a common stable dense domain, where they commute; yet the unitary groups 
they generate do not commute! Hence, it is customary to write 
\eqref{eq:gen_comm_rel} as a more appealing shorthand for the corresponding
regular form  \eqref{eq:gen_weyl_rel}, which however should be fixed without 
ambiguity. Typically, the regular form is understood precisely to be the result
of a formal application of the Baker--Campbell-Hausdorff formula, which we recall it cannot be 
applied in general to unbounded operators. 
In a sense, regular representations are 
precisely those particularly nice representations which 
match with the formula.

These concepts first arose in the famous analysis of 
the uniqueness problem of the canonical commutation relations
\[
[P,Q]=-iI
\]
done by von Neumann, by implementing the ideas of Weyl.
Starting from a physically motivated choice
of the regular representation \cite{weyl}, 
von Neumann \cite{neumann} found a Lie group (the Heisenberg group) reproducing 
precisely the initial regular representations. The study of regular canonical 
representations then was reduced precisely to the representation theory of the
Heisenberg group.


\begin{thebibliography}{99}
\bibitem{lukierski} J.\ Lukierski, A.\ Novicki, H.\ Ruegg and V.\ N.\ Tolstoy,
{\itshape q-Deformation of Poincar\'e algebra}, Phys.\ Lett.\ B 
{\bfseries 268}, 331--338 (1991).
\bibitem{lukierski2} J.\ Lukierski, A.\ Novicki and H.\ Ruegg,
{\itshape New quantum Poincar\'e algebra and \(\kappa\)-deformed
field theory}, Phys.\ Lett.\ B {\bfseries 293}, 344--352 (1992).
\bibitem{majid} S.\ Majid and H.\ Ruegg, {\itshape Bicrossproduct structure of
\(\kappa\)-Poincar\'e group and noncommutative geometry}, Phys.\ Lett.\ B 
{\bfseries 334}, 348--354 (1994).%hep-th/9405107
\bibitem{dp}L.\ D\k abrowksi and G.\ Piacitelli, {\itshape
Canonical Weyl Operators on \(\kappa\)-Minkowski Spacetime}, in preparation.
\bibitem{lukierski_BCH} P.\ Kosinski, J.\ Lukierski and P.\ Maslanka,
{\itshape Local Field Theory on $\kappa$-Minkowski Space, 
Star Products and Noncommutative Translations}, 
Czech.\ J.\ Phys. {\bfseries 50}, 1283 (2000). 
\bibitem{lizzi} A.\ Agostini, F.\ Lizzi and A.\ Zampini, {\itshape
Generalized Weyl Systems and \(\kappa\)-Minkowski space}, Mod.\ Phys.\ Lett.\ A 
{\bfseries 17} 2105-2126 (2002).
\bibitem{lizzi_f} J.\ M.\ Gracia Bond\'{\i}a, F.\ Lizzi, G.\ Marmo and 
P.\ Vitale, {\itshape Infinitely many star products to play with},
JHEP {\bf 0204} 26 (2002). 
\bibitem{neumann}J.\ von Neumann,
         {\itshape Uber die Eindeutigkeit der Schr\"odingerschen Operatoren},
         Math.\ Annalen\ {\bfseries 104}, 570--578 (1931).
\bibitem{agostini} A.\ Agostini, {\itshape \(\kappa\)-Minkowski representations
on Hilbert spaces}, J.\ Math.\ Phys.\ {\bfseries 48}, 052305 (2007). 
\bibitem{rs}M.\ Reed and B.\ Simon, {\itshape Modern Methods of 
Mathematical Physics. I: Functional Analysis}, Academic Press, New York, 1972. 
\bibitem{weyl} H.\ Weyl, {\itshape Gruppentheorie und Quantenmechanik}, Hirzel,
Leipzig, 1928.




\end{thebibliography}
\end{document}